\documentclass[11pt]{article}
\usepackage{moriond,epsfig}

\def\bbn{$\beta \beta 2 \nu$}
\def\bb{$\beta \beta 0 \nu$}
\def\mo{$^{100}$Mo}
\def\se{$^{82}$Se}

\def\be{\begin{equation}}
\def\ee{\end{equation}}
\def\bea{\begin{eqnarray}}
\def\eea{\end{eqnarray}}

\begin{document}
\vspace*{4cm}
\title{Results of the NEMO3 experiment}

\author{J.S Ricol}

\address{CENBG, Chemin du Solarium, Le Haut-Vigneau\\
BP 120, F-33175 Gradignan, France
}

\maketitle

\abstracts{
The purpose of the NEMO3 experiment is to detect neutrinoless double beta decay
in order to determine the nature of neutrino and its absolute mass.
We analysed the 389 effective days of data from
the $\sim 7$ kg of $^{100}$Mo and $\sim 1$ kg of $^{82}$Se and obtained the 
following limits on the half-life for the \bb~process:
$T_{1/2}(\beta\beta0 \nu)~>~4.6~\times~10^{23}$ years (Mo) and 
$T_{1/2}(\beta \beta 0 \nu)~> ~1.0~\times 10^{23}$ years (Se).
The corresponding limits on the neutrino effective mass are 
$\langle m_{\nu}\rangle~<$~0.7~-~2.8~eV~(Mo) and $\langle m_{\nu}>~\rangle$~1.7~-~4.9~eV (Se) at 90\% Confident Level.
We also performed a detailled analysis on the double beta decay of $^{100}$Mo
into the excited states $0^+_1$, $2^+_1$ of $^{100}$Ru.
The results are:
$T_{1/2}(\beta \beta 2 \nu~\rightarrow~0^+_1)~=~5.7^{+1.3}_{-0.9}(stat)\pm~0.7~(syst)~\times~10^{20}$ years,
$T_{1/2}(\beta \beta 2 \nu~\rightarrow~2^+_1)~>~1.1~\times~10^{21}$ years,
$T_{1/2}(\beta \beta 0 \nu~\rightarrow~0^+_1)~>~8.9~\times~10^{22}$ years,
$T_{1/2}(\beta \beta 0 \nu~\rightarrow~2^+_1)~>~1.6~\times~10^{23}$ years.
}

\section{Introduction}

Neutrino physics has encountered a very large success in the last decades.
Neutrino oscillation is an accepted fact and most of the parameters are now measured with a good accuracy.\\
Nevertheless important questions still remain: what is the nature of neutrino, is it a Dirac 
or a Majorana particle ? and what is its absolute mass ?
Double beta decay experiments could solve these 2 questions by detecting the 0 neutrino ($\beta \beta 0 \nu$) mode.
Such a decay would imply the non conservation of the leptonic number and the neutrino to be a Majorana particle.\\
The effective mass of the neutrino is directly related to the  $\beta \beta 0 \nu$ decay rate and
to the nuclear matrix elements (NME) of the involved nuclei transition.
NME are the main uncertainty in this calculation and make difficult the possibility to compare
experiments using different nuclei.
In order to help theorists to contraint their models and converge to a unique solution,
experimentalists need to give them as much data as possible.
The $\beta \beta 2 \nu$ rate and spectra are of course a fruitful source of information but are even more useful 
when compared with the decay into the excited states of the daughter nucleus.
Indeed the behavior of NME parameter is completely different for transition
to the ground and excited states~\cite{gex}.\\
I will first say few words about double beta decay, then describe the NEMO3 detector,
and finally present the main results we obtained recently.

\section{Double beta decay}
The main motivation of double beta decay experiments is the search of the neutrinoless mode :
$$(A,Z) \rightarrow (A,Z+2) + 2 e^-$$
The observation of this decay would imply the leptonic number violation and establish the Majorana nature of neutrino.
The allowed 2 neutrinos mode, already observed for most of double 
beta decay nuclei,
is also a source of interest and need to be studied in detail.\\
Several processes could be responsible for the 0 neutrino mode.
The simpliest and most likely one is the exchange of a light neutrino between the two vertices.
But $\beta \beta 0 \nu$ decay could also be caused by (V+A) currents, Majoron emission or interaction involving 
supersymetric particles.
In any case we can constraint the effective neutrino mass as function of the decay rate.
For example in the case of a light neutrino exchange the formula is:
$$T_{1/2}^{-1} = F(Q_{\beta\beta}^5,Z) |NME|^2 \langle m_\nu \rangle^2,$$
where $F$ is the phase space factor, $NME$ the nuclear matrix elements
and $\langle m_\nu \rangle^2$ the effective mass squared that depends on the mixing matrix parameters and neutrinos mass.\\
To distinguish between all different modes the energy spectrum is a good information
but the angular distribution between the 2 electrons
is also a very powerful tool.

\section{NEMO3 detector}
The NEMO3 (Neutrino Ettore Majorana Observatory) detector~\cite{nim}
is operated in the Frejus underground laboratory since February 2003.
The two main isotopes present in the detector in the form of very thin foils (60 mg/cm$^2$)
are \mo~(6.914 kg, $Q_{\beta \beta} = 3.034$ MeV) and \se~(932 g, $Q_{\beta \beta} = 2.995$ MeV).
On both side of the source there is a gaseous tracking detector which consists of 6180 open drift cells
operating in the Geiger mode and that reconstructs particle track in three dimensions.
Surrounding the tracking detector, the calorimeter, made of 1940 plastic scintillators coupled to
low background PMTs, has an energy resolution (FWHM) of about 14\% at 1 MeV.
The detector is surrounded by a magnetic field that allows the measure the particle charge signature.
The whole detector is covered by two types of shielding against external $\gamma$-rays and neutrons.\\
The NEMO3 detector allows the identification
of electrons, positrons, alphas and gammas, that makes it a unique double beta decay experiment since its a
powerful tool to measure precisely the background
and to constraint different hypothesis for the \bb~decay
using the angular distribution between the two electrons.

\section{Results}

\begin{figure}
\begin{center}
\psfig{figure=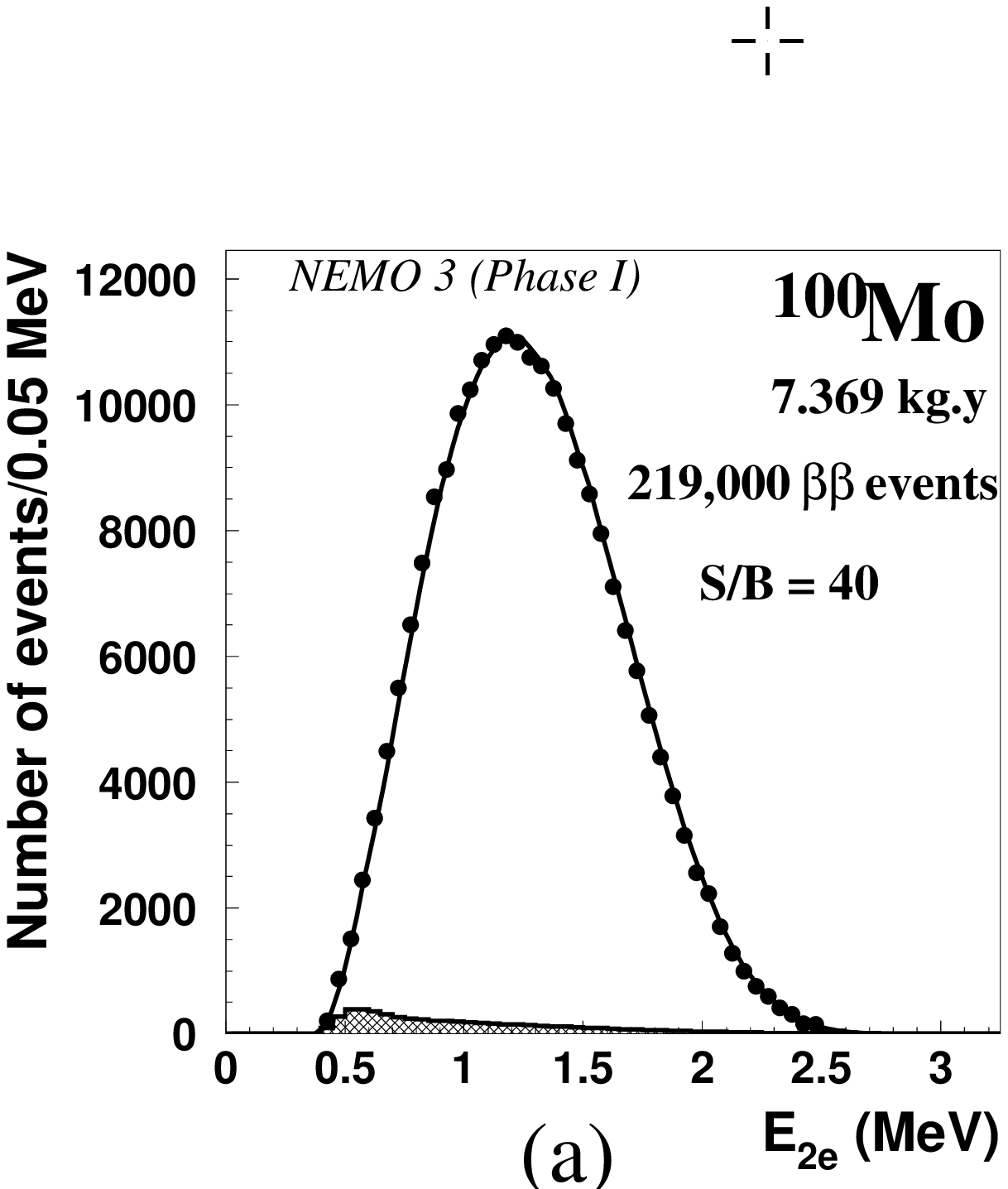,height=2.in}
\hspace*{2.cm}
\psfig{figure=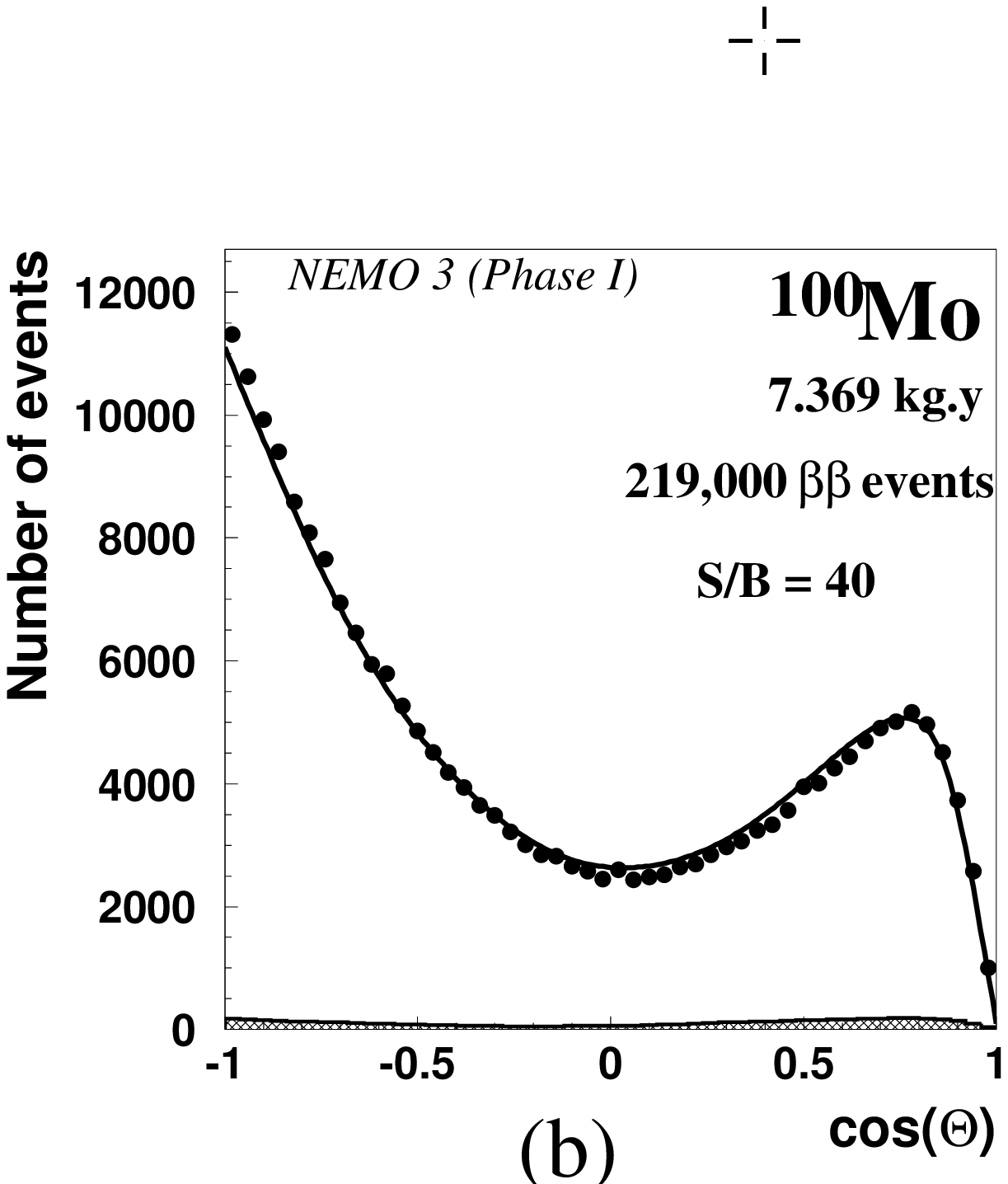,height=2.in}
\caption{Energy sum of the 2 electrons (a) and angular distribution (b) of the \bbn~for \mo.}
\label{fig:2nu}
\end{center}
\end{figure}

The data presented here have been taken between February 2003 and September 2004 (Phase I)~\cite{nemo}.
The double beta events are selected by asking two tracks with negative charge curvature and coming from
a common vertex in the source,
each track being associated to a fired PMT with an energy greater
than 200 keV.
In order to reduce the background we also ask no isolated gamma (no isolated PMT) and no delayed track
(to prevent from Bi-Po coincidence) to be present.\\
The results for \mo~\bbn~decay are presented in Figure~\ref{fig:2nu}
where we can see a very clear signal (S/B = 40).
Both energy and angular distribution are very consistent with Monte Carlo.
We obtained similar results for \se~but with less statistics.
The measured half-life are $T_{1/2}(\beta\beta 0\nu)~=~7.11\pm~0.02~(stat)~\pm~0.54~(syst)~\times~10^{18}$
years for \mo~and
$T_{1/2}(\beta\beta 0\nu)~=~9.6~\pm~0.3~(stat)~\pm~0.54~(syst)~\times~10^{18}$ years for \se.
A maximum likelihood analysis gave the limits on the majoron emission process to be
$T_{1/2}(\beta\beta\chi 0)>2.7~\times~10^{22}$ years.

\subsection{Neutrinoless double beta decay}

\begin{figure}
\psfig{figure=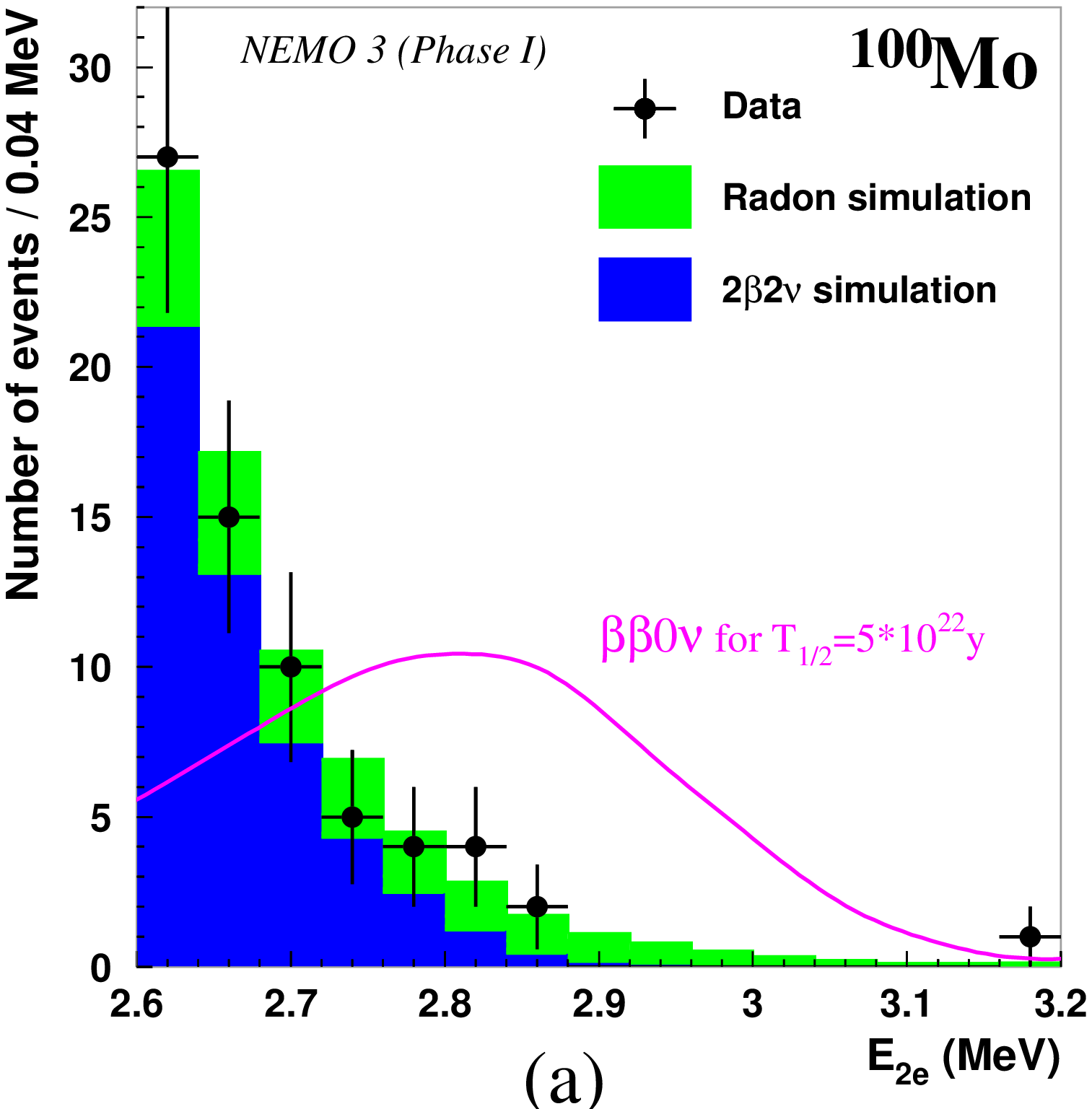,height=2.in}
\hspace*{0.1cm}
\psfig{figure=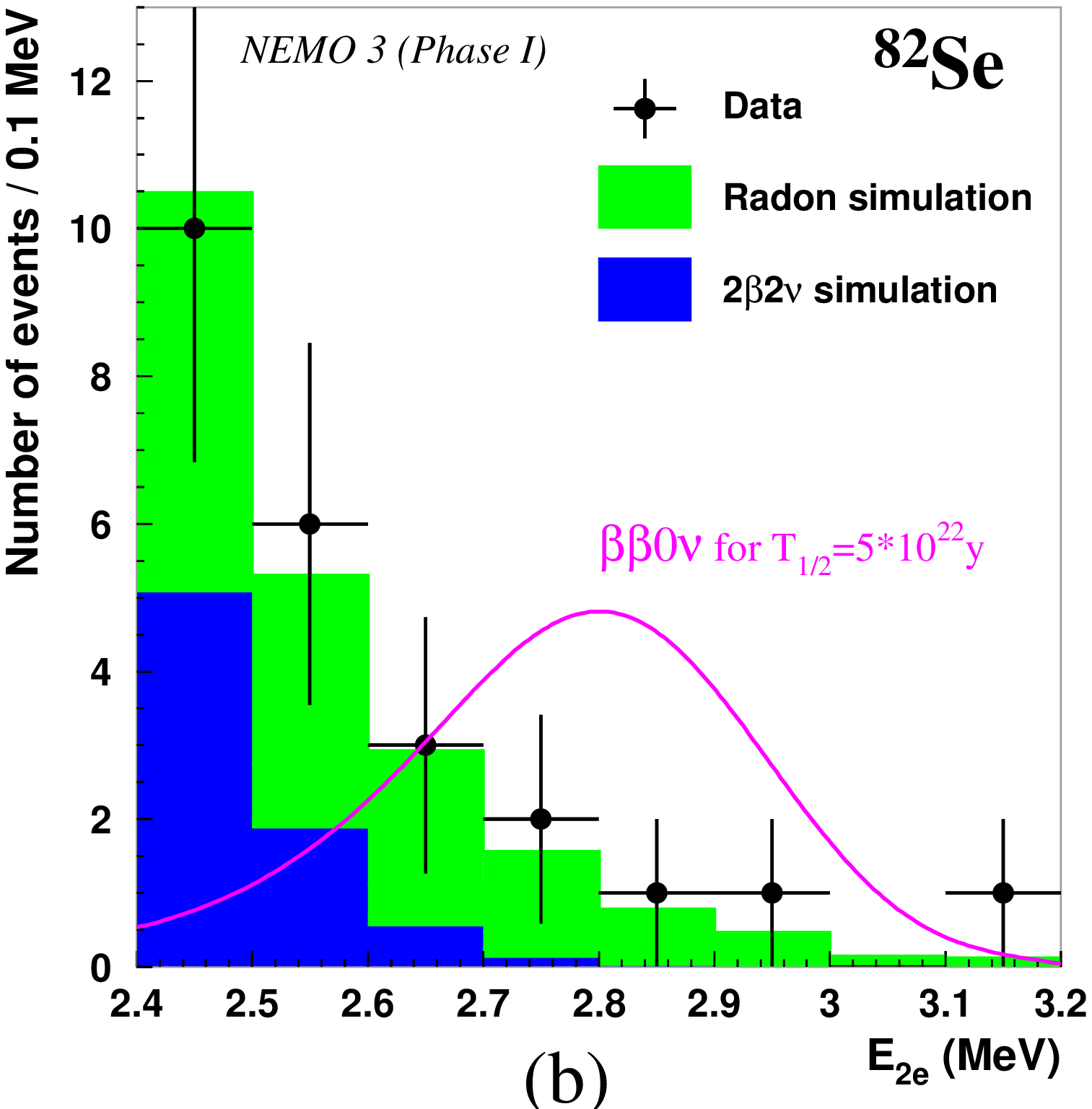,height=2.in}
\hspace*{0.1cm}
\psfig{figure=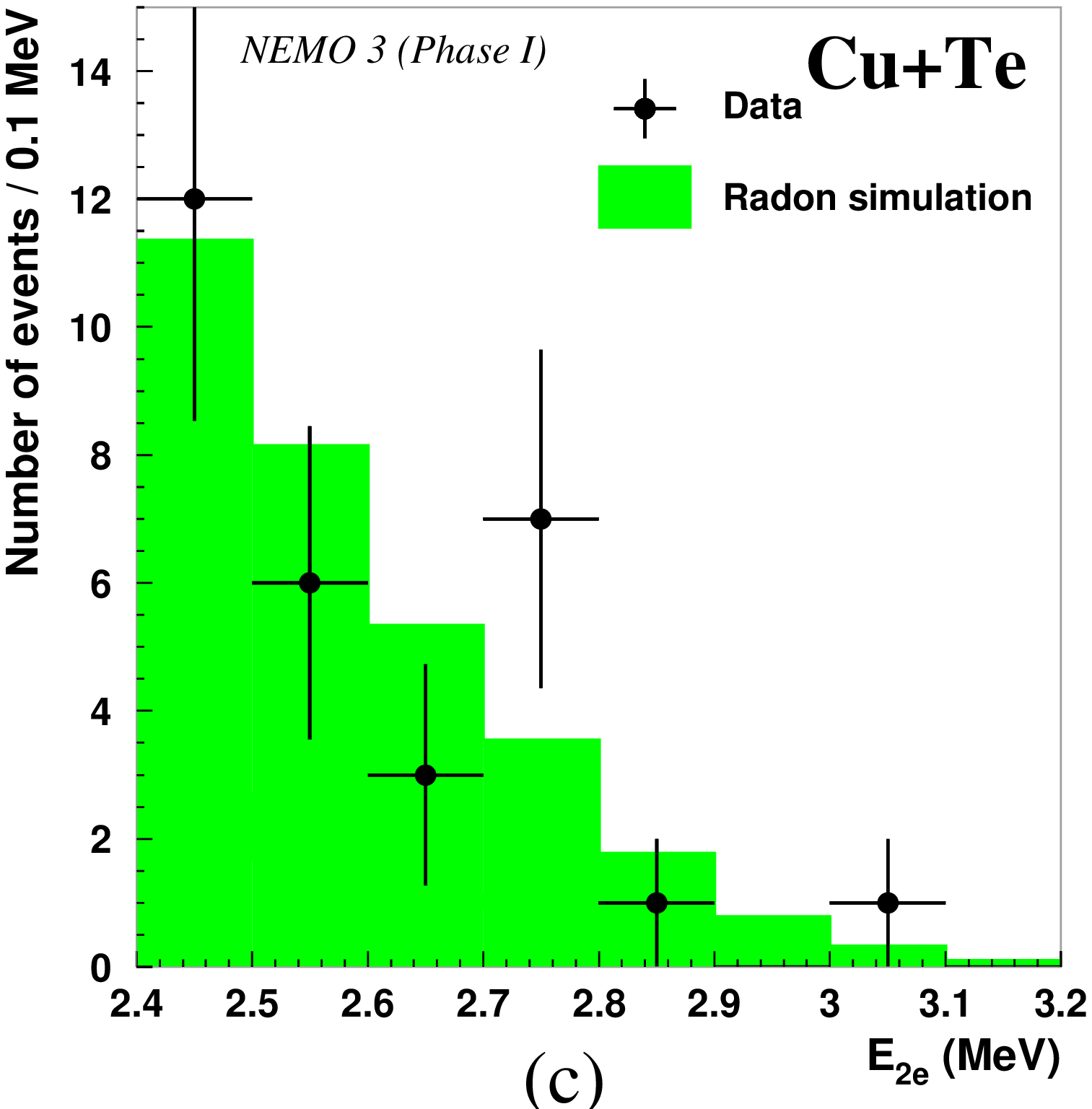,height=2.in}
\caption{Energy spectra in the neutrinoless region for the \mo~(a), \se~(b), and Cu+Te (c).
}
\label{fig:0nu}
\end{figure}

NEMO3 is able to measure each component of the background by different analysis channel:
Compton electrons for external $^{214}$Bi
and $^{208}$Tl, crossing electrons for external neutrons or gammas,
internal ($e^-,\gamma\gamma$) and ($e^-,\gamma\gamma\gamma$) for
$^{208}$Tl impurities in the source,
and ($e^-, delayed~\alpha$) events for $^{222}$Rn that is the dominant background in this
set of data ($25 \pm 5$ mBq/m$^3$).
In the \bb~region, $2.8 < E_1 + E_2 < 3.2$ MeV, the expected number of events is
estimated to be 1.4 events ($\sim~1$~event due to radon, 0.3 from \bbn~and 0.1 from Tl in the foil)
per year and kilogram.
Figure~\ref{fig:0nu} presents the energy spectra in the neutrinoless double beta region.
No evidence for \bb~decay has been found and the data are consistent with the Monte Carlo.
The Cu and Te data (no $\beta \beta$ decay in the selected energy region) are also analysed to study the background,
they show good agreement with radon measurement.
The limits obtained are $T_{1/2}($\bb$) > 4.6 \times 10^{23}$ years, $\langle m_{\nu} \rangle < 0.7 - 2.8$ eV for \mo~and
$T_{1/2}($\bb$) > 1.0 \times 10^{23}$ years, $\langle m_{\nu} \rangle < 0.7 - 2.8$ eV  for \se,
incertainties coming from the choice in the NME.\\
The V+A hypothesis has been analysed and the limit $T_{1/2}($\bb$) > 1.7 \times 10^{23}$ years
has been obtained.\\
NEMO3 is still running and in December 2004, a radon-free air factory has been built which made the Rn
background decrease to the negligible level of 1-2 mBq $/$ m$^3$.
The data in the low-radon period are still under analysis and will lead to more sensitive results.

\subsection{Excited states}

The search of double beta decay to the excited states of $^{100}$Ru is an important study
since it brings additional information for the calculation of NME.
Despite the lower rate of these decays compared to the ground state decay,
this search is helped by the very clear topology of the signal:
in addition to the usual selection cuts for double beta decay
we have one ($2^+_1$ state) or two ($0^+_1$ state) photons in time and with fixed energy.
The kinematic cuts have been optimized from massive Monte Carlo simulations.
After the analysis of 8024 hours of data, the half-life for the \bbn~decay of \mo~to the excited
$0^+_1$ state of $^{100}$Ru is measured to be
$T_{1/2}^{2\nu}(0^+ \rightarrow 0^+_1) = 5.7^{+1.3}_{-0.9} (stat) \pm 0.8 (syst) \times 10^{20}$ years.
Information on energy and angular distributions is also obtained.
Using the phase space value $G = 1.64 \times 10^{-19} y^{-1}$ (for $g_a = 1.254$), we obtain the nuclear
matric element value $M(0_1^+) = 0.103 \pm 0.011$, that is $20\%$ lower than the value to the ground state
$M(g.s) = 0.126 \pm 0.005$.
The search for the double beta decay to the $2^+_1$ state led to the limit
$T_{1/2}^{2\nu}(0^+ \rightarrow 2^+_1) > 1.1 \times 10^{21}$ years,
while the search for neutrinoless double beta decay gave the limit
$T_{1/2}^{0\nu}(0^+ \rightarrow 0^+_1) > 8.9 \times 10^{22}$ years
and $T_{1/2}^{0\nu}(0^+ \rightarrow 2^+_1) > 1.6 \times 10^{23}$ years.

\section{Conclusion}

The NEMO3 detector showed its ability to measure and understand all its background.
No evidence for \bb~has been found in the radon period and the analysis of the data taken in the
low-radon period is in progress and will reach better sensitivity.
We improved the limits on the majoron and V+A decay modes.
Finally we obtained new measurements and improved limits on the decay of the \mo~into excited states.

\end{document}